\documentclass[aps,prl,twocolumn]{revtex4-1}

\usepackage{amsmath}
\usepackage{graphicx}
\usepackage{color}

\begin{document}

\title{How much two-photon exchange is needed to resolve the proton form factor discrepancy?}
\date{\today}
\author{Axel Schmidt}
\email{axelschmidt@gwu.edu}
\affiliation{Massachusetts Institute of Technology, Cambridge, MA 02139, USA}
\affiliation{George Washington University, Washington, DC 20052, USA}

\begin{abstract}
One possible explanation for the proton form factor discrepancy is a contribution to the 
elastic electron-proton cross section from hard two-photon exchange (TPE), a typically
neglected radiative correction. Hard TPE cannot be calculated in a model-independent way,
but it can be determined experimentally by looking for deviations from unity in the ratio 
of positron-proton to electron-proton cross sections. Three recent experiments have measured
this cross section ratio to quantify hard TPE. 
To interpret the results of these experiments, it is germane to ask: 
``How large of a deviation from unity is necessary to fully resolve 
the form factor discrepancy?'' 
With a minimal set of assumptions and using global fits to unpolarized
and polarized elastic scattering data, I estimate the necessary size of the two-photon
exchange correction in the kinematics of the three recent experiments and compare to
their measurements. 
I find wide variation when using different global fits, implying that the magnitude
of the form factor discrepancy is not well-constrained. 
The recent hard TPE measurements can easily accommodate the hypothesis that
TPE underlies the proton form factor discrepancy.
\end{abstract}

\pacs{25.30.Bf 25.30.Hm 13.60.Fz 13.40.Gp }

\keywords{form factor ratio; two-photon exchange; elastic electron scattering; elastic positron scattering}

\maketitle

\section{Introduction}

There is a considerable discrepancy between unpolarized Rosenbluth measurements and
polarized measurements of the proton's electromagnetic form factor ratio,
$R_{FF} \equiv \mu_p G_E / G_M$. Hard two-photon exchange (TPE), a previously neglected
radiative effect, has been suggested as a possible explanation for the discrepancy 
\cite{Guichon:2003qm,Blunden:2003sp}.
While the effect of hard TPE cannot be calculated in a model-independent way, it 
can be determined experimentally, by measuring the deviation
from unity in $R_{2\gamma} \equiv \sigma_{e^+p} / \sigma_{e^-p}$, the 
ratio of the positron-proton to electron-proton elastic cross sections. Three
recent experiments measured $R_{2\gamma}$ over a range of squared 4-momentum transfer, $Q^2$,
up to 2~GeV$^2$ \cite{Rachek:2014fam,Adikaram:2014ykv,Henderson:2016dea} 
(working in units where the speed of light, $c=1$), 
but the results showed only modest hard TPE, leaving open the
question whether or not hard TPE is in fact the cause of the proton form factor
discrepancy. See Ref.~\cite{Afanasev:2017gsk} for a recent review.

To interpret the results of these new experiments, it is helpful to ask the question:
``How much two-photon exchange is needed to resolve the proton form factor discrepancy?''
This question doesn't have a precise answer. One challenge is estimating the size of the
discrepancy itself. There have been dozens of experimental determinations of $R_{FF}$
at many different values of $Q^2$ and these results must be combined, averaged, and 
interpolated. Fortunately, there have been several global fits to both polarized and
unpolarized form factor data, and while they may differ slightly in methodology or
included data, they can provide a parameterization for $R_{FF}$ as determined by
the two different techniques. A second challenge is that the exact kinematic dependence
of the hard TPE effect is unknown. There is not a unique way to translate from the
size of the discrepancy at a given value of $Q^2$ to the necessary value of $R_{2\gamma}$ 
as a function of both $Q^2$ and $\epsilon$, the virtual photon polarization parameter.
Many model-dependent calculations of TPE have been made (see, for example,
Refs.~\cite{Kivel:2012vs,Tomalak:2014sva, Blunden:2017nby}, and others), and these
provide a valuable guide for interpreting experimental results.
However, rather than adopt any model or calculation framework, I propose a method of
estimating the TPE contribution necessary to resolve the form factor discrepancy from
form factor data alone, relying on three reasonable assumptions:
\begin{enumerate}
\item Polarized measurements accurately determine $R_{FF}$, i.e., they are unaffected
by hard TPE. This is the general consensus of the community \cite{Guichon:2003qm,Carlson:2007sp,Arrington:2011dn,Afanasev:2017gsk}.
\item Hard TPE makes no contribution to the elastic cross section in the limit $\epsilon \rightarrow 1$. 
  This is supported by the majority of theoretical calculations of hard TPE (see Refs~\cite{Kivel:2012vs,Tomalak:2014sva, Blunden:2017nby}
as examples, as well as the discussion in \cite{Arrington:2003qk}). 
\item Hard TPE preserves the linearity of Rosenbluth plots. This may not be true, especially
at extreme kinematics, but is very-well supported by previous unpolarized data (and thoroughly
studied in Ref.~\cite{Tvaskis:2005ex}). 
\end{enumerate}
These three assumptions, combined with global fits to unpolarized measurements of $G_E$, 
unpolarized measurements of $G_M$, and polarized measurements of $R_{FF}$ are sufficient
to define the value of $R_{2\gamma}$ that would fully explain the form factor discrepancy.

In this paper I use three different global fits to unpolarized measurements to make predictions
of the hard TPE effect necessary to resolve the form factor discrepancy. I compare these predictions
to the results of the recent TPE experiments, at VEPP-3~\cite{Rachek:2014fam},
at CLAS~\cite{Adikaram:2014ykv,Rimal:2016toz}, and the OLYMPUS Experiment~\cite{Henderson:2016dea}. 
I find that in the $Q^2$ range relevant for these experiments (up to 2~GeV$^2$),
the spread in predictions from using different global fits is very large, indicating that the size of the 
form factor discrepancy is not well constrained. The recent TPE measurements
fall within the spread of predictions, indicating consistency with the hypothesis that TPE is the
origin of the form factor discrepancy. 

\section{Derivation}

To preserve the linearity of Rosenbluth plots, hard TPE must correct the reduced cross section in a way that 
satisfies:
\begin{equation}
G_M^2(Q^2) + \frac{\epsilon}{\tau}G_E^2(Q^2) - \delta(Q^2)(1-\epsilon) = \tilde{G}_M^2(Q^2) + \frac{\epsilon}{\tau}\tilde{G}_E^2(Q^2),
\label{eq:tpe_mod}
\end{equation}
where $G_E$ and $G_M$ represent the true form factors, $\tilde{G}_E$ and $\tilde{G}_M$ represent 
the form factors extracted from unpolarized Rosenbluth separation without accounting for hard TPE, 
$\tau \equiv Q^2/4m_p^2$, where $m_p$ is the proton mass, and $\delta(Q^2)$ represents a lepton charge-odd
modification due to hard TPE. Given the $\epsilon$ dependence of both sides of equation \ref{eq:tpe_mod},
two relationships must hold at every value of $Q^2$:
\begin{align}
G_E^2 &= \tilde{G}_E^2 - \tau \delta\\
G_M^2 &= \tilde{G}_M^2 + \delta.
\end{align}
Dividing the two, one finds that
\begin{equation}
R_{FF}^2 = \frac{\mu_p^2 (\tilde{G}_E^2 - \tau \delta )}{\tilde{G}_M^2 + \delta},
\end{equation}
which can be solved for $\delta$:
\begin{equation}
\delta = \frac{\mu_p^2 \tilde{G}_E^2 - R_{FF}^2 \tilde{G}_M^2}{R_{FF}^2 + \mu_p^2 \tau}.
\end{equation}
By using global fits to unpolarized data to supply $\tilde{G}_E$ and $\tilde{G}_M$ and a global fit
to polarized data to supply $R_{FF}$, an estimate of the value of $R_{2\gamma}$ needed to resolve the 
discrepancy can be made:
\begin{equation}
R_{2\gamma} = 1 + \frac{ 2\delta (1-\epsilon)}{ \tilde{G}_M^2 + \frac{\epsilon}{\tau} \tilde{G}^2_E}.
\label{eq:r2g}
\end{equation}

This approach of estimating $R_{2\gamma}$ from the size of the form factor discrepancy has been employed
by many others in the past starting from a range of assumptions and using a variety of assumed functional forms
\cite{Guichon:2003qm,Arrington:2003qk,Chen:2007ac,Borisyuk:2007re,Guttmann:2010au,Borisyuk:2010ep,Qattan:2011zz,Qattan:2011ke,Bernauer:2013tpr,Qattan:2017dyz}. 
In Ref.~\cite{Borisyuk:2010ep}, Borisyuk and Kobushkin derived an expression that is mathematically equivalent
to that of Eq.~\ref{eq:r2g} though using a slightly different set of assumptions. In Ref.~\cite{Qattan:2011ke},
Qattan et al.\ employ the expression of Ref.~\cite{Borisyuk:2010ep} to extract TPE from several Rosenbluth
separation data sets. In this work, I use global fit models of $\tilde{G}_E$ and $\tilde{G}_M$ to estimate the size of the
TPE correction to resolve the discrepancy for the kinematics of the three recent $R_{2\gamma}$ measurements.

\section{Global Fit Models}

For this method, suitable global fits of $\tilde{G}_E$ and $\tilde{G}_M$ must consider only unpolarized
cross section measurements and not include any hard TPE corrections, either on the cross sections or in
the fit parameterization. Many well known proton form factor
parameterizations (e.g.\ Refs.~\cite{Friedrich:2003iz,Kelly:2004hm,Arrington:2007ux}) are therefore not
suitable. I consider three suitable fits to exclusively unpolarized elastic electron-proton cross sections:
\begin{itemize}
\item Bosted (1995) \cite{Bosted:1994tm},
\item Arrington (2004), unpolarized \cite{Arrington:2003qk},
\item Bernauer et al. (2013), unpolarized \cite{Bernauer:2013tpr}.
\end{itemize}
These fits differ in their parameterization, but more significantly in the input data that are considered.
Bosted fits a representative sample of elastic scattering data, which are described in Ref.~\cite{Walker:1993vj}.
The Arrington fit, whose procedure is described in Ref.~\cite{Arrington:2003df}, includes newer high-$Q^2$ data 
from Jefferson Lab~\cite{Dutta:2003yt,Niculescu:thesis,Christy:2004rc}, as well as additional low-$Q^2$ data from
Mainz~\cite{Borkowski:1974tm, Simon:1980hu} and Saskatchewan~\cite{Murphy:1974zz}. The Bernauer et al.\ fit
includes the 2010 Mainz measurements \cite{Bernauer:2010wm}, comprising approximately 1400 new data points up
to $Q^2=1$~GeV$^2$, in addition to previous world data at larger $Q^2$. For comparison with these global 
fits, I also consider the standard dipole parameterization.
\begin{equation}
G_E(Q^2) \approx \frac{1}{\mu_p}G_M(Q^2) \approx \left(1 + \frac{Q^2}{0.71 \text{GeV}^2} \right)^{-2}.
\end{equation}

A suitable global fit for $R_{FF}$ should consider only polarization measurements, without any incorporation
of TPE-corrected unpolarized cross section measurements. This type of fit has not yet been of 
significant interest so no extremely sophisticated fits of this kind have been published. Gayou 
et al.\ perform a linear fit in the range of $0.5 < Q^2 < 5.6$~GeV$^2$ \cite{Gayou:2001qd}. 
For this paper, I will also use a linear model that is consistent with the world polarization data:
\begin{itemize}
\item $R_{FF}(Q^2) = 1 - (0.12 \text{~GeV}^{-2}) Q^2$.
\end{itemize}

Bernauer et al.\ conveniently report uncertainty estimates on $\tilde{G}_E$, $\tilde{G}_M$, and
their ratio, allowing me to make an uncertainty estimate on the extracted $R_{2\gamma}$. My estimate is
approximate; the uncertainties on $\tilde{G}_E$ and $\tilde{G}_M$ are correlated, and these correlations
are not reported. Except at very low $Q^2$ however, the uncertainty on $\tilde{G}_{E}$ is dominant. Therefore, 
my uncertainty estimates on $R_{2\gamma}$ extracted using the Bernauer et al.\ fits are based on the uncertainty
on $\tilde{G}_E$ only. I am further neglecting any contribution to the uncertainty from my linear
model of $R_{FF}$, making my uncertainty estimates an underestimate.
It should be noted that, since the Bosted and Arrington fits are based on fewer data than the
Bernauer fits, one would expect them to have uncertainties that are at least as large as those of
Bernauer et al., and significantly larger for $Q^2 < 1$~GeV$^2$.

\begin{figure}[htpb]
\centering
\includegraphics[width=\columnwidth]{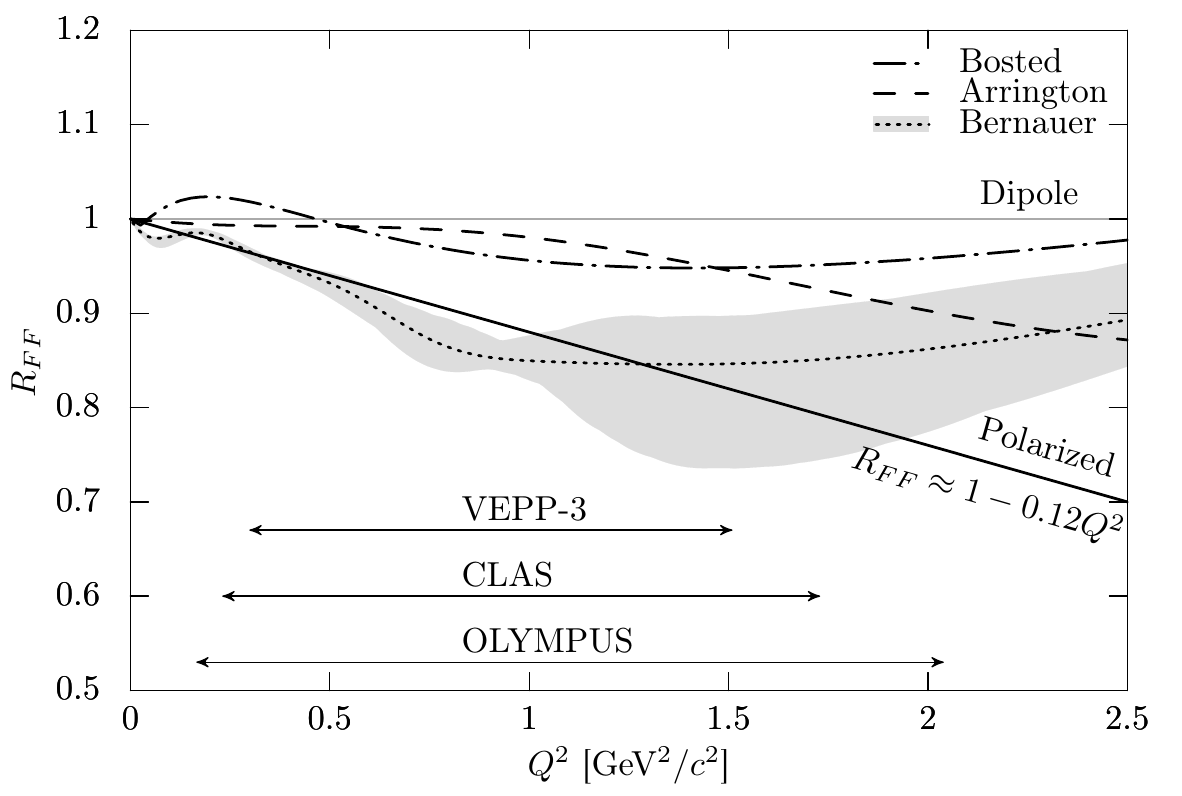}
\caption{\label{fig:RFF} The proton form factor ratio, $R_{FF}$, is shown as a function of $Q^2$ for the
fits employed in this work. The Bernauer et al.\ unpolarized fit predicts a significantly smaller form factor
discrepancy than the fits of Arrington and Bosted. The $Q^2$ coverage of the three recent TPE experiments is
shown with arrows.}
\end{figure}

Fig.~\ref{fig:RFF} shows $R_{FF}$ as predicted by the three unpolarized global fits as a function of $Q^2$,
as well as the $Q^2$ coverage of the three recent TPE experiments. The proton form factor discrepancy is
essentially the deviation between the unpolarized and polarized predictions of $R_{FF}$. The size of the
discrepancy varies considerably between the different unpolarized fits.

One remark must be made regarding the consistency of radiative corrections. Bosted, Arrington, and Bernauer et 
al.\ make explicit efforts to make sure that consistent radiative corrections were re-applied to all input cross
sections before fitting. However, they chose to apply different correction prescriptions. Bosted and Arrington
follow a prescription based on Mo and Tsai~\cite{Mo:1968cg}, 
with some improvements detailed in the appendices of Ref.~\cite{Walker:1993vj}. By contrast, Bernauer et al.\ adopted the
prescription by Maximon and Tjon~\cite{Maximon:2000hm} and further exponentiated the correction to account
for radiation at all orders. In addition, Bernauer et al., also chose to apply a Feshbach correction for Coulomb 
distortion. The exact choice of radiative corrections prescription may seem like a small technical detail,
but it will alter the form factors obtained from a fit (see, for example, Ref.~\cite{Arrington:2007ux}).
The choice of prescription amounts to an assertion of what corrections are necessary to make the measured reduced
cross sections linear in $\epsilon$ for fixed $Q^2$. However, without redoing the fits, it is difficult to assess
the magnitude of the effect on $G_E/G_M$---since these corrections are non-linear in $\epsilon$, the effect
depends on the $\epsilon$ coverage of the input cross section data. Therefore, I make no attempting to correct
the Bernauer et al.\ extractions of the form factors to unify approaches with Arrington and Bosted. I use the
global fits as published and the results should be interpreted with this caveat in mind.

The choice of radiative correction prescriptions also affects the interpretation of measurements of $R_{2\gamma}$,
since, for example, Maximon and Tjon use a different definition of soft TPE than do Mo and Tsai. The $R_{2\gamma}$
data shown in this work all use the Mo and Tsai definition (OLYMPUS has published results for multiple prescriptions
~\cite{Henderson:2016dea}).

\section{Results}

\begin{figure}[htpb]
\centering
\includegraphics[width=\columnwidth]{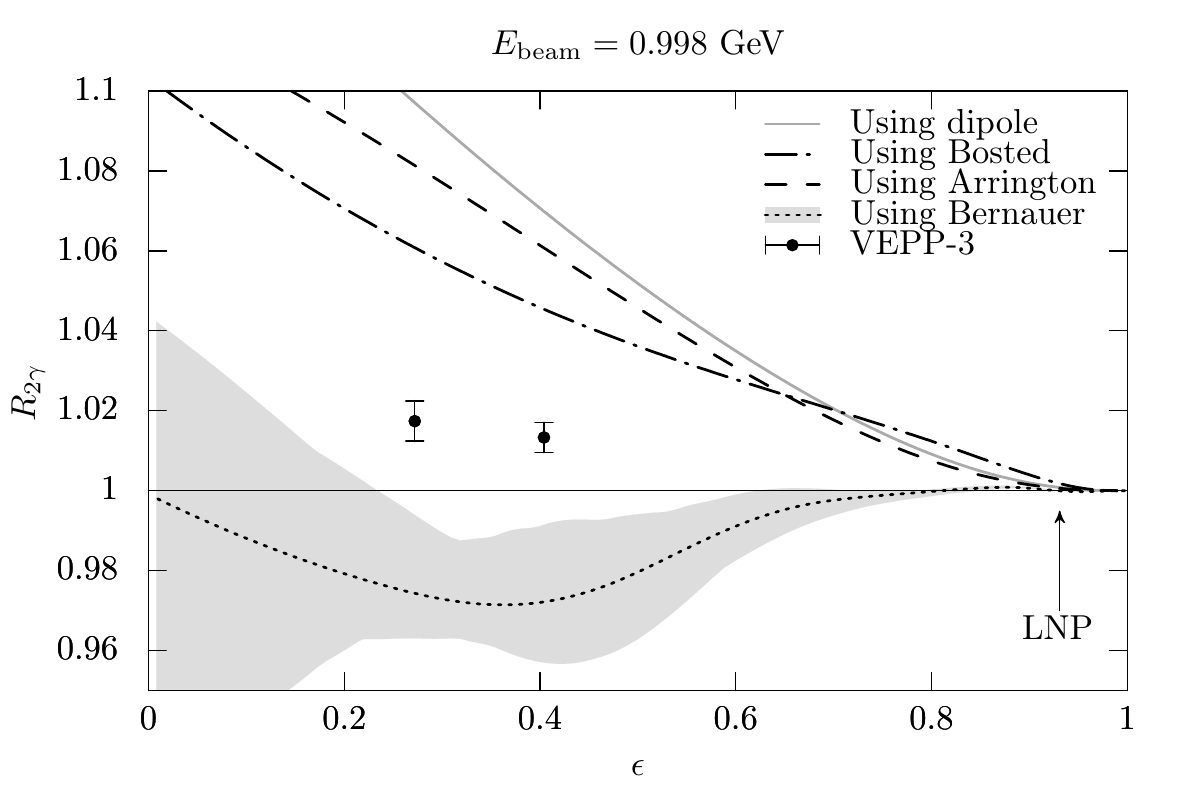}\\
\includegraphics[width=\columnwidth]{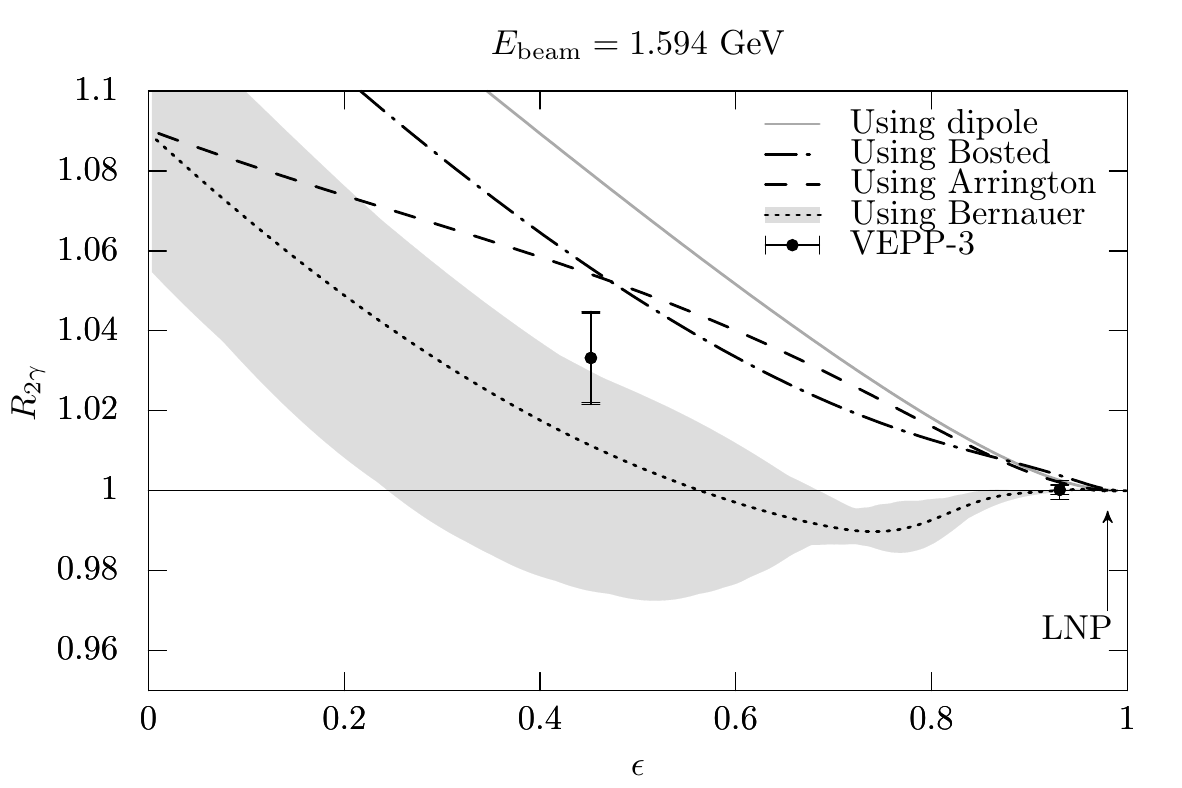}
\caption{\label{fig:vepp} The results from the VEPP-3 TPE experiment~\cite{Rachek:2014fam} for a beam energy
of 0.998~GeV (top panel) and 1.594~GeV (bottom panel) fall below predictions based
the Bosted and Arrington fits but above the prediction based on the Bernauer fit.}
\end{figure}

\begin{figure}[htpb]
\centering
\includegraphics[width=\columnwidth]{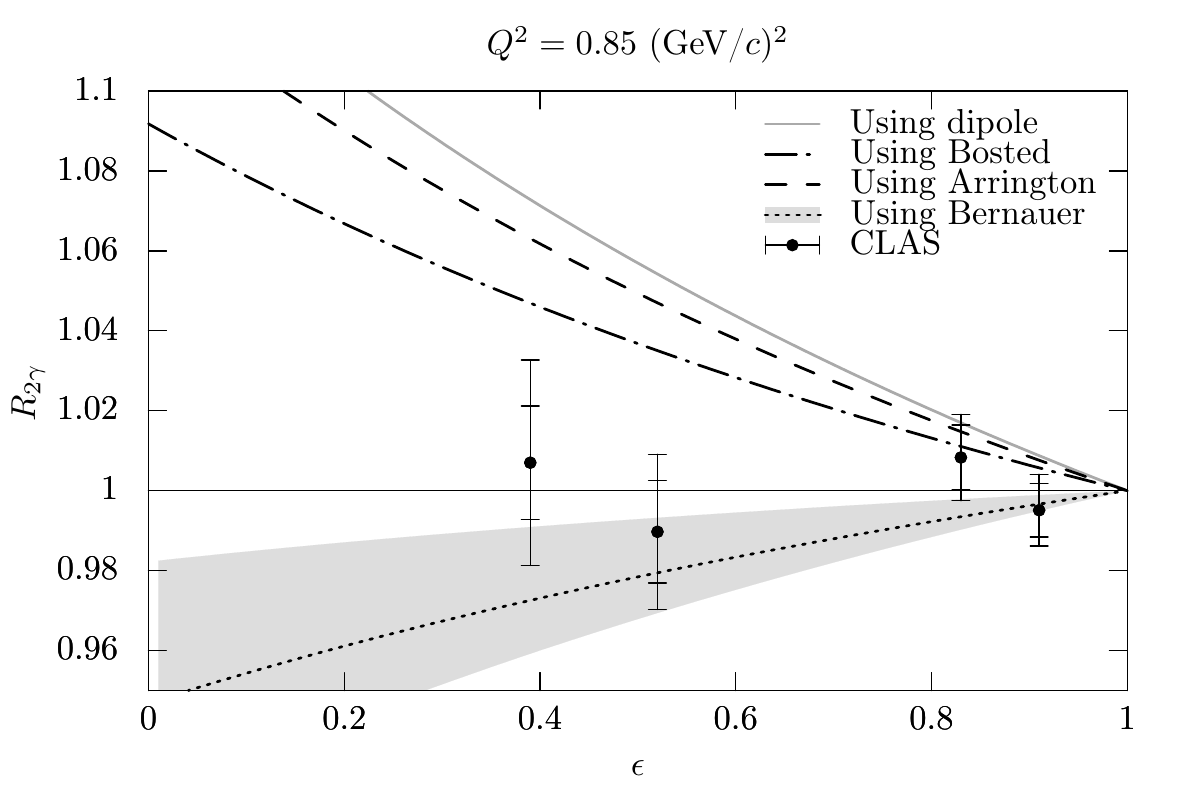}\\
\includegraphics[width=\columnwidth]{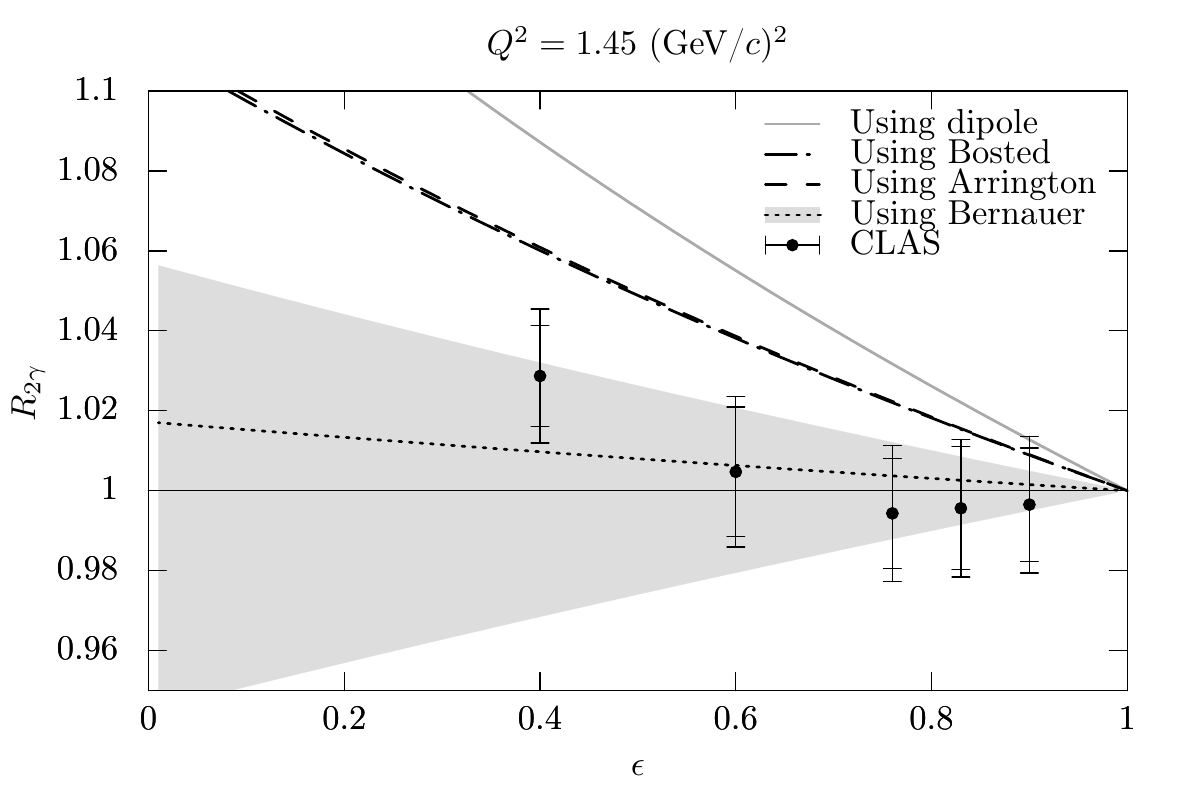}
\caption{\label{fig:clas} The results from the CLAS TPE experiment~\cite{Rimal:2016toz} for $Q^2=0.85$~GeV$^2$ (top panel)
and $Q^2=1.45$~GeV$^2$ also fall below the prediction produced using the Bosted and Arrington fits,
but above that coming from the Bernauer fit.}
\end{figure}

\begin{figure}[htpb]
\centering
\includegraphics[width=\columnwidth]{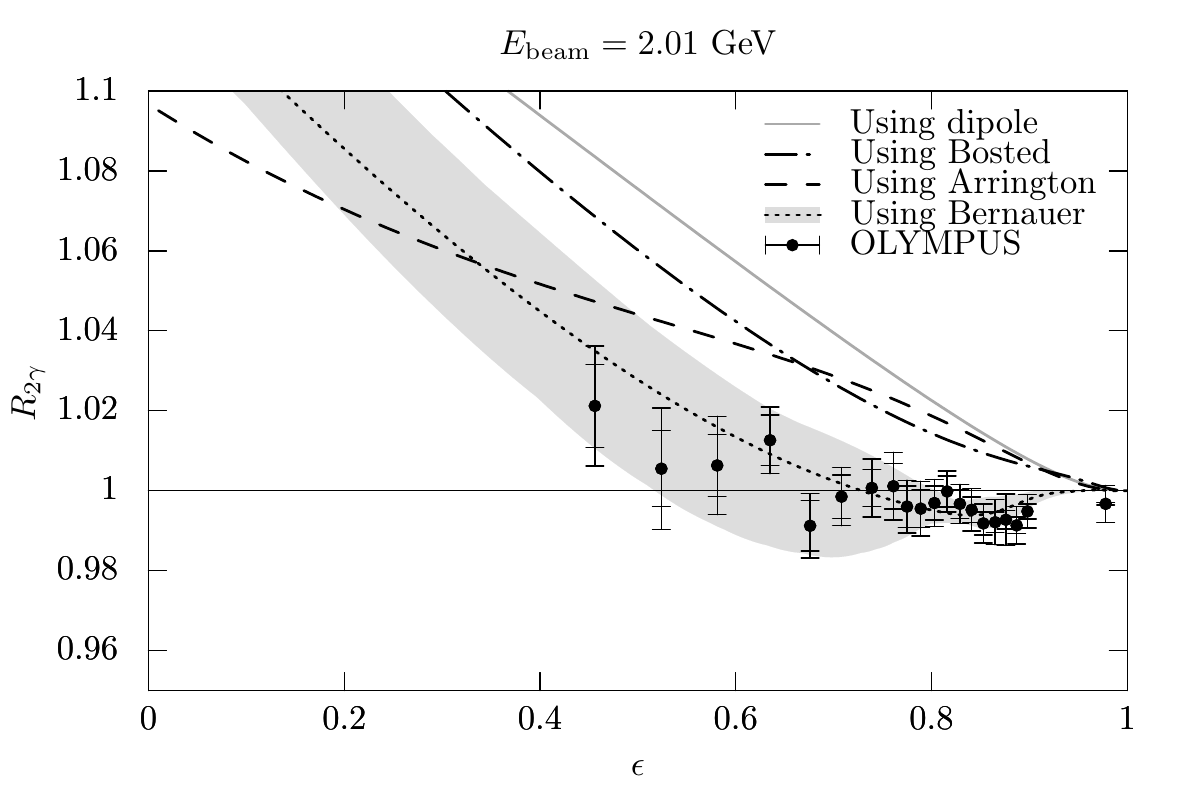}
\caption{\label{fig:olympus} OLYMPUS results~\cite{Henderson:2016dea} are close to the prediction
based on the Bernauer fit, but have a smaller slope.}
\end{figure}

Figs.~\ref{fig:vepp}--\ref{fig:olympus} show predictions for $R_{2\gamma}$ based on 
Eq.~\ref{eq:r2g} as functions of $\epsilon$ in the kinematics of the VEPP-3, CLAS, 
and OLYMPUS two-photon exchange experiments, compared with their respective results.
As a general trend, the measured data fall within or slightly above the uncertainty
band using the Bernauer fits, but below the predictions using the Bosted and Arrington fits.

The results of the two runs of the VEPP-3 two-photon exchange experiment are shown
in Fig.~\ref{fig:vepp}. The inner error bars show the statistical uncertainty, while
the outer error bars show the statistical and systematic uncertainties added in quadrature.
Arrows mark the luminosity normalization points (LNPs), 
the kinematic point to which $R_{2\gamma}$ was normalized. In comparing the data to predictions,
the measured values of $R_{2\gamma}$ can float relative to the value of $R_{2\gamma}$
at the LNP. The band associated with the prediction using Bernauer et al.\ indicates
the uncertainty arising from the statistical and systematic uncertainty (added in quadrature) on $\tilde{G}_E$.
The data from both beam energies show an increasing $R_{2\gamma}$ with 
decreasing $\epsilon$, which is the correct sign for explaining the discrepancy.
The magnitude of this increase falls between the prediction of the Bernauer fits
and those of the Bosted and Arrington fits. 

The results from the CLAS TPE experiment, using their constant $Q^2$ binning scheme~\cite{Rimal:2016toz},
are shown in Fig.~\ref{fig:clas}. Inner error bars show the statistical uncertainty, while outer error bars
show the statistical and systematic uncertainties added in quadrature. A normalization uncertainty of 0.003
is not shown. Like with the VEPP-3 results, the CLAS data are below the predictions using the Bosted 
and Arrington fits, but are reasonable consistent with those using the Bernauer fits when accounting for
uncertainties.

The results from the OLYMPUS experiment~\cite{Henderson:2016dea}, with exponentiated Mo and Tsai
radiative corrections, are presented in Fig.~\ref{fig:olympus}. The inner error bars show statistical
uncertainty, while the outer error bars show statistical and point-to-point systematic uncertainties
added in quadrature. Additional correlated uncertainty ranging from 0.0036 to 0.0045 is not shown. 

The OLYMPUS results have a non-zero slope, increasing with decreasing $\epsilon$, indicating a hard
TPE contribution. However, at high epsilon, the data fall below $R_{2\gamma}=1$. The OLYMPUS results
are closest to the prediction based on the Bernauer fit, but with less slope. Meanwhile, the predictions
based on the Bosted and Arrington are significantly above the OLYMPUS data.

\section{Discussion}

There are two general trends that can be seen in Figs.~\ref{fig:vepp}, \ref{fig:clas},
and \ref{fig:olympus}. First, the prediction based on the Bernauer fits is significantly different
from those based on Bosted, Arrington, and the standard dipole, even given the uncertainty estimate.
The inclusion of the 2010 Mainz data
set has a large effect on the apparent size of the form factor discrepancy. Looking at Fig.~\ref{fig:RFF},
Bernauer shows no discrepancy up to $Q^2 \approx 1.3$~GeV$^2$. The $R_{FF}$ difference between the
Bernauer fits and the others are driven largely by the differences in $G_M$ at low $Q^2$. This 
suggests that as long as there is a lack of consensus on $G_M$, there will be uncertainty on how big the proton
form factor discrepancy actually is, and on how much TPE is needed to resolve it. More unpolarized
cross section data, especially at low $Q^2$ and backward angles, would provide valuable constraints on $G_M$. 
New results, such as those from the PRad Experiment~\cite{Xiong:2019umf} will at least allow updated
global form factor fits that may help solidify the situation.

Second, the recent TPE data fall below the predictions using the Bosted and Arrington fits, but above
the prediction using the Bernauer fits. If the Bosted and Arrington form factor fits are to be believed,
the data do not support the hypothesis that TPE is the sole cause of the form factor discrepancy. Judging
by the Bernauer fit prediction, there is adequate TPE. The data so far cannot make any definitive claims,
and can easily accommodate the TPE hypothesis. As is clear from Fig.~\ref{fig:RFF}, higher $Q^2$ data are
needed for a more definitive test.

Given both the spread in predictions based on different form factor fits, and the large uncertainties
indicated by the Bernauer fits, it is clear that a proper and comprehensive uncertainty analysis is
needed. Such an analysis must take into account the correlations between fit parameters, the 
correlations they introduce between $G_E$ and $G_M$, and the resulting uncertainty on $R_{2\gamma}$. 

As new elastic electron-proton scattering data become available, the technique I describe can be used
to improve our understanding of the proton form factor discrepancy and the amount of hard TPE needed
to resolve it. This can provide valuable context for the interpretation of upcoming experiments, such
as MUSE~\cite{Gilman:2017hdr}, and those being considered at DESY and Mainz.

\begin{acknowledgments}
I thank J.~C.~Bernauer, R.~M.~Milner, and D.~K.~Hasell for their helpful advice on this work.

This work was supported by the Office of Nuclear Physics of the U.S. Department of Energy, grant No. DE-FG02-94ER40818.

\end{acknowledgments}

\appendix

\bibliography{references}

\end{document}